\documentclass[twocolumn,prl,showpacs,amsfonts,amsmath,assymb,eufrak]{revtex4}

\usepackage{epsfig}
\usepackage{color}
\usepackage{bm}

\newcommand{\todayd}{\the\year/\the\month/\the\day}

\newcommand{\bib}{\bibitem}

\newcommand{\bel}{\begin{easylist}}
\newcommand{\eel}{\end{easylist}}
\newcommand{\bi}[1]{\begin{itemize} #1 \end{itemize}}
\newcommand{\be}[1]{\begin{enumerate} #1 \end{enumerate}}

\def \({\left(}
\def \){\right)}

\newcommand{\sumtwo}[2]%
{\mathop{\sum_{#1}}_{#2}}
\newcommand{\sumthree}[3]%
{\mathop{\mathop{\sum_{#1}}_{#2}}_{#3}}
\newcommand{\sumfour}[4]%
{\mathop{\mathop{\mathop{\sum_{#1}}_{#2}}_{#3}}_{#4}} 
\newcommand{\prodtwo}[2]%
{\mathop{\prod_{#1}}_{#2}}
\newcommand{\mintwo}[2]%
{\mathop{\min_{#1}}_{#2}}
\newcommand{\maxtwo}[2]%
{\mathop{\max_{#1}}_{#2}}
\newcommand{\maxthree}[3]%
{\mathop{\mathop{\max_{#1}}_{#2}}_{#3}}
\newcommand{\limtwo}[2]%
{\mathop{\lim_{#1}}_{#2}}
\newcommand{\suptwo}[2]%
{\mathop{\sup_{#1}}_{#2}}
\newcommand{\supthree}[3]%
{\mathop{\mathop{\sup_{#1}}_{#2}}_{#3}}
\newcommand{\supfour}[4]%
{\mathop{\mathop{\mathop{\sup_{#1}}_{#2}}_{#3}}_{#4}} 
\newcommand{\inftwo}[2]%
{\mathop{\inf_{#1}}_{#2}}
\newcommand{\infthree}[3]%
{\mathop{\mathop{\inf_{#1}}_{#2}}_{#3}}
\newcommand{\inffour}[4]%
{\mathop{\mathop{\mathop{\inf_{#1}}_{#2}}_{#3}}_{#4}} 

\def\rnum#1{\resizebox{0.5em}{\height}{\expandafter{\romannumeral #1}}}
\def\Rnum#1{\resizebox{0.5em}{\height}{\uppercase\expandafter{\romannumeral #1}}}

\newcommand{\Mea}{Mondaini {\it et al}. }

\begin{document}

\preprint{APS/123-QED}

\title{Shiraishi and Mori Reply}

\author{Naoto Shiraishi}
\affiliation{%
Department of Physics, Keio University, 3-14-1 Hiyoshi, Yokohama, Japan
}%

\author{Takashi Mori}%
\affiliation{Department of Physics, The University of Tokyo,7-3-1 Hongo, Bunkyo-ku, Tokyo, Japan}

\date{\today}

\maketitle

The preceding Comment by \Mea\cite{comment} on our letter~\cite{SM} raised two issues.
One is on the notion of the eigenstate thermalization hypothesis (ETH) and the other is on our inappropriate description in the introduction.
As for the second issue, we agree with this comment (this is our misdescription~\cite{misdesc}).
In this reply, we concentrate on the first issue, and argue that our formulation of the ETH is a standard and natural one in the context of thermalization, and that our results are qualitatively new and nontrivial.

Let us start by distinguishing the {\it diagonal ETH} and the {\it off-diagonal ETH} for a given macroscopic observable.
The diagonal ETH has an established definition that the eigenstate expectation values are described by a smooth function of energy.
In contrast, there exist several definitions of the off-diagonal ETH~\cite{Sre96,AGH,Gol,DAl} (see \cite{offETH-com}).
Although the diagonal ETH and the off-diagonal ETH naturally appear simultaneously in the random matrix and quantum chaos theory~\cite{Sre96,DAl}, these two have completely different roles and properties in the context of thermalization.
In this context, the word ``ETH" often means the diagonal ETH~\cite{KIH,BMH,Pal,GE16,Tas16,Bir,Sor,Khe,Sir,Rig08}, and our letter~\cite{SM} follows this convention.

We now go back to the Comment.
\Mea consider a system with no local conserved quantity (LCQ) but with nonlocal conserved quantities associated with some symmetries.
Correspondingly, the Hilbert space is divided into several symmetry sectors.
Throughout this Reply, we treat this type of systems~\cite{LCQ-def}.
The authors' interest is in the ETH in the whole Hilbert space (i.e., all sectors) and that in each sector.
We label them as
\vspace{5pt}

(d1): the diagonal ETH in the whole Hilbert space,

(d2): the diagonal ETH in each sector,

(o1): the off-diagonal ETH in the whole Hilbert space,

(o2): the off-diagonal ETH in each sector.
\vspace{5pt}

\noindent The Comment numerically shows that both (d1) and (o2) hold, but (o1) is violated.
On the basis of this, \Mea claim that the ETH should be considered in each sector and the violation of (d1) in systems with no LCQ is not surprising.
We agree that the property of the {\it off-diagonal} ETH depends on the specific sector.
However, the argument below asserts that the {\it diagonal} ETH should be considered in the whole Hilbert space in the context of thermalization, and that the violation of {(d1)}, which is the main claim of our letter~\cite{SM}, is nontrivial and unexpected before us.

We now argue that the roles of the diagonal and off-diagonal ETH in thermalization are completely different, and there are good reasons to put emphasis on (d1).
An initial state thermalizes if and only if
\be{
\item The long-time average of a macroscopic observable is equal to that of microcanonical ensemble,

\vspace{-3pt}

\item Its time-series fluctuation is small.
}
The diagonal ETH concerns (i) and the off-diagonal ETH concerns (ii).
In fact, the off-diagonal ETH ensures (ii)~\cite{DAI, vio-offETH}.
However, (ii) is also ensured by largeness of the effective dimension $D_{\rm eff}$ of the initial state.
The time-series fluctuation is proven to be bounded above by of order $O(1/\sqrt{D_{\rm eff}})$~\cite{Rei08,LPSW,SF12}.
It is numerically shown~\cite{Rig16} and theoretically proven~\cite{FBC16} that a generic initial state has a sufficiently large effective dimension.
Thus, (ii) generically realizes even without the off-diagonal ETH, and therefore the main subject in the field of thermalization has been whether (i) holds.

We emphasize that the diagonal ETH in the sense of (d1) guarantees (i)~\cite{GE16, Tas16}, while (d2) does not~\cite{d2-thermal}.
Therefore, if one wants to derive thermalization via ETH, the ETH should be interpreted as (d1), against the interpretation by \Mea.
Importantly, previous numerical studies~\cite{SR2,KIH,BMH,Sor}, including \Mea\cite{comment}, have shown that (d1) holds in systems with no LCQ.

Taking these background into consideration, our results~\cite{SM, MS}
\bi{
\item A model with no LCQ violates (d1)

\vspace{-3pt}
\item This model without (d1) thermalizes after quench
}
are highly nontrivial and important for the research of thermalization.
We emphasize that contrary to \Mea\cite{comment} the generic violation of (o1) never implies the violation of (d1), because generic systems with no LCQ violate (o1) but satisfy (d1)~\cite{generic}.
Their numerical result on (o1), which might be interesting by itself, is irrelevant to our letter~\cite{SM}.

\end{document}